\documentclass[12pt]{article}
\usepackage{scicite}

\usepackage{times}

\usepackage{graphicx}

\newenvironment{sciabstract}{%
\begin{quote} \bf}
{\end{quote}}

\newcounter{lastnote}

\newcommand{\gtrsim}{\,\stackrel{>}{\sim}\,}

\title{Optical Images of an Exosolar Planet 25 Light Years from Earth$^\ast$}

\author
{Paul Kalas$^{1\ast}$,  James R. Graham$^{1}$, Eugene Chiang$^{1,2}$, Michael P. Fitzgerald$^{3}$, \\
Mark Clampin$^{4}$,
Edwin S. Kite$^{2}$,  Karl Stapelfeldt$^{5}$, 
Christian Marois$^{6}$, \\
John Krist$^{5}$
\\
\\
\normalsize{$^{1}$Astronomy Department, University of California,}\\
\normalsize{Berkeley, CA 94720, USA}\\
\normalsize{$^{2}$Department of Earth \& Planetary Science, University of California, Berkeley, CA 94720, USA}\\
\normalsize{$^{3}$Institute of Geophysics \& Planetary Science, Lawrence Livermore National Laboratory}\\
\normalsize{Livermore, CA 94551, USA}\\
\normalsize{$^{4}$Exoplanets \& Stellar Astrophysics Laboratory, Goddard Space Flight Center,} \\
\normalsize{Greenbelt, MD 20771, USA}\\
\normalsize{$^{5}$MS 183-900, Jet Propulsion Laboratory, Caltech, Pasadena, CA 91109, USA}\\
\normalsize{$^{6}$Herzberg Institute for Astrophysics, Victoria, BC V9E2E7, CANADA}\\
\\
\normalsize{$^\ast$Accepted for publication in {\it Science}}
}

\date{}

\begin{document}

\baselineskip24pt

\maketitle

\begin{sciabstract}
Fomalhaut is a bright star 7.7 parsecs (25 light years) from Earth
that harbors a belt of cold dust with a structure consistent with
gravitational sculpting by an orbiting planet.  Here, we present
optical observations of an exoplanet candidate, Fomalhaut b.  In the
plane of the belt, Fomalhaut b lies approximately 119 astronomical
units (AU) from the star and 18 AU from the dust belt, matching
predictions.  We detect counterclockwise orbital motion using Hubble
Space Telescope observations separated by 1.73 years.  Dynamical
models of the interaction between the planet and the belt indicate
that the planet's mass is at most three times that of Jupiter for the
belt to avoid gravitational disruption.  The flux detected at 0.8
$\mu$m is also consistent with that of a planet with mass no greater
than a few times that of Jupiter.  The brightness at 0.6 $\mu$m and
the lack of detection at longer wavelengths suggest that the detected
flux may include starlight reflected off a circumplanetary disk, with
dimension comparable to the orbits of the Galilean satellites.  We
also observed variability of unknown origin at 0.6 $\mu$m.
\end{sciabstract}

\section*{}

Approximately 15\% of nearby stars are surrounded by
smaller bodies that produce copious amounts of fine dust via
collisional erosion ({\it 1}).  These ``dusty debris
disks'' are analogues to our Kuiper Belt, and can be imaged directly through
the starlight they reflect or thermal emission from their dust grains.
Debris disks may be gravitationally sculpted by more massive
objects;  their structure gives indirect evidence for the existence
of accompanying planets (e.g., {\it 2, 3}).  
Fomalhaut, an A3V star 7.69 pc from the Sun ({\it 4}), is an excellent example:
a planet can explain both the observed 15 AU offset between the
star and the geometric center of the belt, and the sharp 
truncation of the belt's inner edge ({\it 3, 5--7}).  With an estimated age
of 100--300 Myr ({\it 8}), any
planet around Fomalhaut would still be radiating its 
formation heat, and would be amenable to direct detection.
The main observational challenge is that Fomalhaut is one of the
brightest stars in the sky ($m_V$=1.2 mag); to detect a planet
around it requires the use of specialized
techniques such as coronagraphy to artificially eclipse the star and
suppress scattered and diffracted light. 

\vspace{0.5cm}
{\noindent \bf Detection of Fomalhaut b}

Coronagraphic observations with the Hubble Space Telescope (HST) in
2004 produced the first optical image of Fomalhaut's dust belt, and
detected several faint sources near Fomalhaut ({\it 6}).  Fomalhaut's
proper motion across the sky is 0.425 arcsecond per year in the
southeast direction, which means that objects that are in the
background will appear to move northwest relative to the star.  To
find common proper motion candidate sources, we observed Fomalhaut
using the Keck II 10-m telescope in 2005 and with HST in 2006 (SOM).
In May, 2008, a comprehensive data analysis revealed that Fomalhaut b
is physically associated with the star and displays orbital motion.
Follow-up observations were then conducted at Gemini Observatory at
3.8 $\mu$m (SOM).

Fomalhaut b was confirmed as a real astrophysical object in six
independent HST observations at two optical wavelengths (0.6 $\mu$m
and 0.8 $\mu$m; see Fig. 1 and Table S1).  It is co-moving with
Fomalhaut, except for a 0.184 $\pm$ 0.022 arcsecond (1.41 $\pm$ 0.17
AU) offset between 2004 and 2006 ($\Delta$T = 1.73 yr), corresponding
to 0.82 $\pm$ 0.10 AU yr$^{-1}$ projected motion relative to Fomalhaut
(SOM).  If Fomalhaut b has an orbit that is coplanar and nested within
the dust belt, then its semimajor axis is $a \simeq$ 115 AU, close to
that predicted by Quillen ({\it 7}). 
%
%
An object with $a$ = 115 AU in Keplerian motion around a star with
mass 2.0 M$_\odot$ has an orbital period of 872 years, and a circular
speed of 3.9 km s$^{-1}$.  The six Keplerian orbital elements are
unconstrained by measurements at only two epochs; however, by
comparing the deprojected space velocity ($5.5 _{-0.7}^{+1.1}$ km
s$^{-1}$) with the circular speed we find a lower limit on the
eccentricity of 0.13 (95\% confidence) by assuming that Fomalhaut b is
at periastron.  Thus, our observations are consistent with bound
Keplerian motion, although the exact range of allowable eccentricity
depends sensitively on poorly known uncertainties in orbital
inclination, apsidal orientation, and host stellar mass.

Fomalhaut b is located near the faint half of the belt seen in stellar
light backscattered by dust grains.  Therefore it lies behind the sky
plane (the Earth-Fomalhaut-Fomalhaut b angle is 126$^{\rm o}$), at
approximately 51$^{\rm o}$ past conjunction as it orbits
counter-clockwise.  Though faint, Fomalhaut b is still one hundred
times brighter than reflected light from a Jupiter-like planet at that
radius from Fomalhaut (SOM).

\vspace{0.5cm}
\noindent {\bf Dynamical Models of Planet-Belt Interaction}

We constrain the mass of Fomalhaut b by modeling its gravitational
influence on the dust belt, reproducing properties of the belt
inferred from the HST scattered light images. Our model assumes that
Fomalhaut b is solely responsible for the observed belt
morphology. This assumption implies that the orbits of the belt and of
Fomalhaut b are apsidally aligned. The deprojected space velocity of
Fom b is nominally inconsistent with this expectation.  Apsidal
misalignment may imply the existence of additional perturbers; then
the mass estimates derived from our single planet models are upper limits.

Our modeling procedure takes four steps. First, for a given mass and orbit of
Fomalhaut b, we create a population of several thousand parent bodies
stable to gravitational perturbations from the planet. These parent
bodies, modeled as test particles, do not undergo close encounters
with Fomalhaut b over 100 Myr.  Initial parent body orbits have
semimajor axes between 120 and 140 AU, and eccentricities and
longitudes of periastron that are purely secularly forced by the
planet ({\it 9}).  Initial inclinations of parent
bodies are randomly and uniformly distributed within 0.025 radian of
Fomalhaut b's orbital plane, and remaining orbital angles are drawn at
random. After 100 My, parent body orbits differ somewhat from these
initial conditions; most survivors have semimajor axes $>$ 130 AU.
The forced orbits thus constructed are nested ellipses of
eccentricity $\approx 0.11$ that approximate the observed belt
morphology. Forced orbits are expected to result from interparticle
collisions, which dissipate random motions and compel planetesimals to
conform towards closed, non-intersecting paths ({\it 10}).

This elliptical annulus of parent bodies is termed a ``birth ring''
({\it 11}); erosive collisions among parent bodies give birth to
smaller sized but more numerous dust grains.  The observed scattered
stellar light arises predominantly not from parent bodies but rather
from their dust progeny.  Thus the second step of our procedure is to
track dust trajectories.  We take each parent to release a dust grain
with the same instantaneous position and velocity as its parent's.
The trajectory of a grain of given $\beta$ (force of radiation
pressure relative to that of stellar gravity; $\beta$ scales inversely
as grain radius) is then integrated forward under the effects of
radiation pressure and Poynting-Robertson drag. We carry out
integrations for $\beta \in (0, 0.00625, 0.0125, \ldots, 0.4)$.  For
$\beta$ approaching the radiation blow-out value of $\sim$1/2, grains
execute highly elongated orbits whose periastra are rooted within the
birth ring.  Integrations last 0.1 Myr, corresponding to the
collisional lifetime of grains in Fomalhaut's belt, as estimated from
the inferred optical depth of the belt.

Third, we superpose the various $\beta$-integrations to construct maps
of optical depth normal to the belt plane.  To reduce the shot noise
associated with a finite number of grains, we smear each grain along
its orbit: each grain is replaced by an elliptical wire whose
linear density along any segment is proportional to the time a
particle in Keplerian motion spends traversing that segment.  We
compute the optical depth presented by the collection of wires,
weighting each $\beta$-integration according to a Dohnanyi ({\it 12})
grain size distribution. This distribution, which reflects a
quasi-steady collisional cascade in which parent bodies grind down to
grains so small they are expelled by radiation pressure, is assumed to
hold in the birth ring, where dust densities are greatest and collision
rates highest.

The final step is to compare the optical depth profile of our
dynamical model with that of a scattered light model adjusted to fit
the 2004 HST image of Fomalhaut's belt ({\it 6}). We focus
on the one belt property that seems most diagnostic of planet mass and
orbit: the belt's inner edge, having a semimajor axis of $a_{\rm
  inner} = 133$ AU according to the scattered light model. This edge
marks the outer boundary of the planet's chaotic zone ({\it 7}).
The chaotic zone is a swath of space enclosing the planet's orbit
which is purged of material because of dynamical instabilities caused
by overlapping first-order mean-motion resonances ({\it 13}). For a
given planet mass $M$, we adjust the planet's semimajor axis $a$ until
the dynamical model's optical depth attains half its maximum value at
$a_{\rm inner}$ (Fig. 2, bottom panel).  Applying this procedure, we find that
\begin{equation}
a_{\rm inner} - a = 2.0 (M/M_{\ast})^{2/7} a
\end{equation}
where $M_{\ast}$ is the central stellar mass.

Two trends that emerge from our modeling
imply that the mass of the planet should be low.  First, as $M$
increases, the planet more readily perturbs dust grains onto eccentric
orbits, and the resultant optical depth profile becomes too broad at
distances $\gtrsim$ 140 AU (Fig.~2, bottom panel).  Second, to not
disrupt the belt, larger mass planets must have smaller orbits,
violating our estimate for the current stellocentric distance of
Fomalhaut b (Fig.~2, top two panels).  Together, these considerations
imply that $M < 3 M_{\rm J}$.
This upper limit supersedes those derived previously ({\it 7}), as the
quantitative details of our model are more realistic [see also ({\it
  14})]: the belt as a whole is modeled, not just its inner edge;
parent bodies are handled separately from dust grains, and only the
latter are used to compare with observations; stellar radiation
pressure is accounted for; parent bodies are screened for dynamical
stability over the system age; and grain-grain collisions are
recognized as destructive, so that dust particle integrations are
halted after a collision time.

\vspace{0.5cm}
{\noindent \bf Model Planet Atmospheres}

Comparison between our photometric data and model planet atmosphere spectra indicate
that Fomalhaut b may be a cooling Jovian-mass exoplanet with age 100-300 Myr (Fig. 3).
A planet atmosphere model with effective temperature T$_{\rm eff}$ = 400 K and radius 1.2 $R_J$,
for which the bolometric luminosity is $3.4\times10^{-7}~L_\odot$ 
({\it 15-16}), reproduces the observed 0.8 $\mu$m flux.
This model implies that the luminosity of Fomalhaut b is lower than any other object observed outside the
solar system, and thus that it is not a young brown dwarf or a more massive object.
Theoretical cooling tracks of objects with $T_{\rm eff}$= 400 K  and ages  $>$100 Myr 
are insensitive to uncertain initial conditions (see figure 1 of {\it 15}).  
The luminosity on these tracks is given by 
$L \sim 2 \times10^{-7} (M / 2M_J)^{1.87} (t / 200 {\rm Myr})^{-1.21} L_\odot$,
implying that the mass of Fomalhaut b is
1.7--3.5 $M_J$.   The error in the mass is 
dominated by the age uncertainty.  

Relative to the models of planet atmospheres, the flux of Fomalhaut b is
too faint by at least a factor of a few at 1.6 $\mu$m, and the upper limit
set by observations at 3.8 $\mu$m is only
marginally consistent with the models.  However, the various models
disagree with each other by similar factors at 1.6 $\mu$m, partly because of theoretical uncertainties
associated with the strengths of the CH$_4$ vibrational bands.
Moreover, our hypothesized effective temperature is near the
condensation temperature of water clouds, and such clouds are a large source
of uncertainty in planet atmosphere models.  Nevertheless, our observations
at 1.6 $\mu$m and 3.8 $\mu$m exclude a warmer (more massive)
planet.

Choosing a 400 K, 46 m s$^{-2}$, 5$\times$ solar abundance model from
({\it 15}) as a baseline, we can investigate the effects of gravity
and composition using theoretical exoplanet model spectra ({\it
  15-16}).  The elevated abundance set is chosen to be representative
of solar system gas giants.  The temperature and gravity of this model
are a good match to a 200 Myr, 2.5 $M_{\rm J}$ exoplanet.  As
previously noted, this model accounts for the 0.8 $\mu$m flux, but
over predicts the 1.6 $\mu$m band flux by a factor of three. Cooler
models (350 K) cannot simultaneously reproduce the 0.8 $\mu$m flux
without violating the long wavelength flux limits, while for hotter
models (500 K) the 1.6 $\mu$m upper limit becomes particularly
problematic.  If there is a significant thermal photospheric
contribution to the 0.8 $\mu$m flux, then 400 K is a rough upper limit
to the temperature of the object.

The 400 K, solar abundance model has reduced methane opacity which
causes it to be unacceptably bright in the $H$ band. The colors and
fluxes also depend on the surface gravity. Models from ({\it 15}) for
10 m s$^{-2}$ and 215 m s$^{-2}$ are also available: the colors of the low
gravity model are too red in both [0.8 $\mu$m $-$ 1.6 $\mu$m] and [0.8
  $\mu$m $-$ 3.8 $\mu$m] to be acceptable.  Thus, if the gravity is
lower than our nominal assumption, corresponding to approximately a
0.5 $M_{\rm J}$ object, then we estimate that the upper limit on
temperature is raised by about 50 K. The colors of the high-gravity
400 K model are similar to those of the 46 m s$^{-2}$ one.

\vspace{0.5cm}
{\noindent \bf Other Sources of Optical Emission}

From 0.6 to 0.8 $\mu$m, Fomalhaut b is bluer than the
models predict (Fig. 3).  Furthermore,
between 2004 and 2006 Fomalhaut b became fainter by $\sim$0.5 mag at
0.6 $\mu$m .  Photometric variability and excess optical emission cannot be explained by
exoplanet thermal radiation alone.  The 0.6 $\mu$m flux might be contaminated  by
H$\alpha$ emission (SOM) that is
detected from brown dwarfs ({\it 17, 18}).  
Variable H$\alpha$ emission might arise from a hot planetary
chromosphere heated by vigorous internal convection, or
trace hot gas at the inner boundary of a circumplanetary
accretion disk, by analogy with magnetospheric emission from 
accreting T Tauri stars (e.g., {\it 19}).
If a circumplanetary disk is extended, the
starlight it reflects might contribute to the flux detected at 0.6 and
0.8 $\mu$m.  To explain
our observed fluxes  requires a disk
radius $\sim20-40$ R$_J$, comparable to the orbital radii of Jupiter's
Galilean satellites (SOM).   The need for additional sources of luminosity implies
that the mass inferred from the 0.8 $\mu$m flux alone is an upper
limit.

As remarkably distant as Fomalhaut b is from its star, the planet
might have formed {\it in situ}. The dust belt of Fomalhaut contains
three Earth masses of solids in its largest collisional
parent bodies.  Adding enough gas to bring this material to cosmic
composition would imply a minimum primordial disk mass of 1 M$_{\rm
  J}$, comparable to the upper mass limit of Fomalhaut b.
Alternatively, the planet might have migrated outward by interacting
with its parent disk ({\it 20}), or by gravitationally scattering off
another planet in the system and having its eccentricity mildly damped
by dynamical friction with surrounding disk material ({\it 21}).  \\

{\bf References and Notes}

\begin{enumerate}
\item D. E. Backman, F. C. Gillett, in {\it Cool Stars, Stellar Systems and
the Sun}, eds. J. L Linsky and R.E. Stencel (Springer-Verlag, Berlin),
pp. 340-350 (1987).
\item D. Mouillet, J. D. Larwood, J.C.B. Papaloizou, A. M. Lagrange, {\it Mon. Not. R. Astron. Soc.} {\bf 292}, 896 (1997).
\item M.C. Wyatt, {\it et al. Astrophys. J.} {\bf 527}, 918 (1999).
\item 1 pc = 3.09$\times$10$^{18}$ cm
\item K. Stapelfeldt, {\it et al., Astrophys. J. Suppl. Ser.} {\bf 154}, 458 (2004).
\item P. Kalas, J. R. Graham, M. Clampin, {\it Nature} {\bf 435}, 1067 (2005).
\item A. Quillen, {\it Mon. Not. R. Astron. Soc.} {\bf 372}, L14 (2006).
\item D. Barrado y Navascues, {\it Astron. Astrophys.} {\bf 339}, 839 (1998).
\item C. D. Murray, S. F. Dermott, {\it Solar System Dynamics.} (Cambridge Univ. Press, Cambridge, U.K., 1999).
\item B. Paczynski, {\it Astrophys. J.} {\bf 216}, 822 (1977).
\item L. E. Strubbe, E. I. Chiang, {\it Astrophys. J.} {\bf 648}, 652 (2006).
\item J. W. Dohnanyi, {\it J. Geophys. Res.} {\bf 74}, 2531 (1969).
\item J. Wisdom, {\it Astron. J.} {\bf 85}, 1122 (1980).
\item E. Chiang, E. Kite, P. Kalas, J R. Graham, M. Clampin, {\it Astrophys. J.}, in press (2008).
\item J.J. Fortney,  {\it et al.,  Astrophys. J.}  {\bf 683}, 1104 (2008).
\item A. Burrows, D. Sudarsky, J. I. Lunine, {\it Astrophys J.} {\bf 596}, 587 (2003).
 \item A. J. Burgasser,  {\it et al., Astron. J.} {\bf 120}, 473 (2000).
 \item C. Marois, B. Macintosh, T. Barman, {\it Astrophys. J.} {\bf 654}, L151 (2007).
 \item L. Hartmann, R. Hewett, N. Calvet, {\it Astron. J.} {\bf 426}, 669 (1994).
\item D. Veras, P. J. Armitage, {\it Mon. Not. R. Astron. Soc.} {\bf 347}, 613 (2004).
\item E. B. Ford, E. I. Chiang, {\it Astrophys. J.} {\bf 661}, 602 (2007).
\item J. Davis, {\it et al.,  Astron. Nachr.}  {\bf 326}, 25 (2005).
\item P.K. acknowledges support from GO-10598, and K.S. and J.K. acknowledge support
from GO-10539, provided by  NASA through a grant from STScI
under NASA contract NAS5-26555.   E.C. acknowledges support from NSF grant AST-0507805.  
M.F. acknowledges support from the Michelson Fellowship Program, 
under contract with JPL, funded by NASA. Work at LLNL was 
performed under the auspices of DOE under contract DE-AC52-07NA27344.
E.K. acknowledges support from a Berkeley
Fellowship.
We thank the staff at STScI, Keck and Gemini for supporting our
observations.
\end{enumerate}

{\noindent\bf Supporting Online Material (SOM)}
\\SOM Text\\
Fig. S1\\
Tables S1 to S4\\
References

\section*{Supporting Online Material}

{\bf Observing Method}\\

Observations with the Hubble Space Telescope (HST) were obtained with
the Advanced Camera for Surveys (ACS) High Resolution Channel (HRC) in
its coronagraphic mode ({\it S1}).  The HRC is a 1024$\times$1024
pixel CCD with a 1.8$^{\prime\prime}$ coronagraphic occulting spot
near the center of the detector, and a 3.0$^{\prime\prime}$ occulting
spot toward the upper left edge.  After a correction for geometric
distortion the pixel size is 25$\times$25 mas.  In 2004 we placed
Fomalhaut behind the 1.8$^{\prime\prime}$ occulting spot only, whereas
in 2006 we imaged Fomalhaut behind both occulting spots (Table S1).
Even though the occulting spots block the core of the stellar point
spread function (PSF), a significant halo of light is present in the
entire CCD frame.  We use two separate strategies to remove this PSF
halo: 1) We observe another bright star (Vega) with the coronagraph
and use this template PSF to subtract the PSF of Fomalhaut, and (2) We
image Fomalhaut such that the detector is rotated at different angles
relative to the sky.  In the instrument reference frame the PSF is
quasi-static, whereas astrophysical features rotate.  The 2006 data
acquire Fomalhaut at four separate position angles (PA) of the sky on
the detector, with a maximum PA separation of 6$^{\rm o}$.  Taking the
median value of these frames gives a master PSF that does not contain
the astrophysical features.  The master PSF is then subtracted from
the individual images, which are then rotated to a common orientation
and combined.  Technique 2 is known in the literature as roll
deconvolution or angular difference imaging (ADI; {\it S2-S4}).

Table S1 catalogs our observations.  Fomalhaut b is detected
independently in each row with an F606W and F814W observations.  For
each of these rows, Fomalhaut b is detected using both PSF subtraction techniques
outlined above.  False-positives are defined as apparent point sources
that cannot be consistently confirmed among these data sets.

\vspace{0.5cm}
\begin{tabular}{lllllllllllll}
\multicolumn{5}{c}{\bf Table S1: Fomalhaut Observing log} \\
\hline
Observatory & Instrument & UT Date & Filter & Exp. Time (s)\\ 
\hline
HST 2.4-m & ACS/HRC 1.8$^{\prime\prime}$ spot & 2004 Oct. 25 	& F606W & 1320\\
		& ................. 1.8$^{\prime\prime}$ spot  & 2004 Oct. 26 	& F606W & 1320\\	
		& ................. 1.8$^{\prime\prime}$ spot & 2006 July 14	& F435W &  6525\\ 
 		& ................. 3.0$^{\prime\prime}$ spot & 2006 July 15-16	& F435W &  6525\\ 				
		& ................. 1.8$^{\prime\prime}$ spot & 2006 July 17-19	& F606W &  7240\\ 
 		& ................. 3.0$^{\prime\prime}$ spot & 2006 July 19-20	& F606W &  7240\\ 
		& ................. 1.8$^{\prime\prime}$ spot & 2006 July 18	& F814W &  5430\\ 
 		& ................. 3.0$^{\prime\prime}$ spot & 2006 July 19 	& F814W &  5430\\ 

Keck II 10-m & NIRC2 2.0$^{\prime\prime}$ spot	& 2005 July 17 & H & 3790\\ 
		 &		& 2005 July 27 & H & 4320\\ 
		 &		& 2005 July 28 & H & 4890\\ 
		 &		& 2005 Oct. 21 & H & 5310\\ 
		 &		& 2005 Oct. 22 & CH4 & 4774\\ 
Gemini N 8-m & NIRI & 2008 Sep. 17-18 & L$^\prime$ & 6006\\

\hline
\end{tabular}
\vspace{0.5cm}

Keck II observations with adaptive optics used the NIRC2 near-infrared camera located at the Nasmyth of the telescope
where the sky rotates relative to the instrument focal plane.  We used a camera scale of
0.04$^{\prime\prime}$ per pixel and a 2.0$^{\prime\prime}$ diameter, semi-transparent occulting spot.
Though the instrument has reimaging optics to fix the sky angle relative
to the detector reference frame, we permit the sky to rotate in
order to employ PSF subtraction technique 2 (ADI).  Gemini South observations at L$'$, without
adaptive optics correction,
were executed in a similar manner to employ the ADI technique.
We used the NIRI F/32 camera with 22 mas pixels, giving a 
$22.4^{\prime\prime}\times22.4^{\prime\prime}$ field of view.
Fomalhaut b is not detected in either the Keck II or Gemini North data.\\

{\noindent\bf Astrometry}

The astrometric reference frame is established relative to the star Fomalhaut
as there are no other adequately bright stars contained within the ACS/HRC
field of view.  The significant source of astrometric uncertainty is determining
the position of Fomalhaut behind the ACS/HRC occulting spots.  Successive
frames may be registered at the sub-pixel level relative to each other by mutual subtraction,
but a fiducial frame is required where the pixel position of Fomalhaut behind
the occulting spot is estimated.  This is achieved by minimizing the 
residuals when a frame is subtracted from a copy 
of itself rotated by 180$^{\rm o}$.  The residuals are minimized when
the assumed center of rotation is nearest the position of the
star behind the spot.  The self-subtraction
center positions can be compared to the relative center positions
determined by subtracting images of Fomalhaut obtained in successive
orbits.  We thus estimate the accuracy of determining the
location of Fomalhaut behind the occulting spots using 
180$^{\rm o}$ self-subtraction technique as  $\pm$0.5 pixel (12.5 mas,
or 0.10 AU at the distance to Fomalhaut).   This value is an upper 
limit to the possible difference between the true and estimated
positions of Fomalhaut.  

The centroid position of Fomalhaut b were measured in three versions
of the final F606W processed images in 2004, and seven versions from the
2006 F606W processed data.  All images were rotated to the orientation
shown in Fig. 1.  We find a standard error of 0.31 and 0.55 pixel
along the x and y directions in the 2004 data.  In the 2006 measurements
the corresponding standard errors are 0.09 and 0.32 pixel. Adding these
uncertainties in quadrature to the uncertainties
in the position of Fomalhaut at each epoch
gives 0.87 pixel for the 1-$\sigma$ uncertainty
in the estimated motion of Fomalhaut b between epochs.
This translates to 0.022 arcsecond or 0.169 AU.  

The fact that Fomalhaut b is orbiting Fomalhaut is robust
because the apparent orbital motion of 7.3 pixels between epochs is significantly
greater than these uncertainties, as well as
the PSF full-width at half-maximum of $\sim$2.7 pixels.  Fomalhaut b
cannot be a background objects as shown in Fig. S1.   
The empirical RMS accuracy in the position angle achieved 
in ACS data is 0.003 degrees  ({\it S5}),  which 
corresponds to an insignificant uncertainty of 0.03 pixel at the radial position 
of Fomalhaut b.
\\

{\noindent\bf Photometry}

In the cases where Fomalhaut b was detected (HST), we report photometry corrected to
an infinite aperture using DAOPHOT and an empirical curve of growth derived
from the data (Table S2).  Zeropoints for the HST data are obtained from ({\it S6}).
The error bars quoted are statistical only.  The standard error derived 
from multiple versions of the F606W data with different PSF subtraction
techniques is 0.10 mag and 0.05 mag for the 2004 and 2006 data, respectively.

\vspace{0.5cm}
\begin{tabular}{lllllllllllll}
\multicolumn{4}{c}{\bf Table S2:  Photometry on Fomalhaut b} \\
\hline
UT Date & Filter & $\lambda_c$ ($\mu$m) &Magnitude & Error (mag) & Detection?\\ 
\hline
2004-10-25 & F606W & 0.606 &24.43 & 0.08 & Yes\\
2004-10-26 & ... & ... &24.29 & 0.09 & Yes\\
2005-07-21 & H  & 1.633 & $>$22.9  & 3$\sigma$ limit & No\\
2005-10-21 & CH$_4$S & 1.592   &$>$20.6  & 3$\sigma$ limit & No\\
2006-07-14/20 & F606W &0.606 & 25.13 & 0.09 & Yes\\
		...	& F814W & 0.814 & 24.55 & 0.13& Yes\\
		...	& F435W & 0.435 & $>$24.7 & 3$\sigma$ limit& No\\
2008-09-17/18 & L$^\prime$  & 3.78& $>$16.6 & 3$\sigma$ limit& No\\
\hline
\end{tabular}
\vspace{0.5cm}
\\

Photometric calibration of the Keck upper limits is a multi-step process.
Data were scaled to a common signal level using background star observations prior to combination.  
For the multi-night combination of $H$-band data, we use the July 17 observations for photometric 
calibration because of that night's exceptional conditions.
The peak brightness of Fomalhaut was measured through the partially transmissive occulting spot in short exposure images.  
These measurements were used to determine an on-axis sensitivity calibration 
using the 2MASS photometry of Fomalhaut and the previously measured occulting 
spot transmission.
We derived a sensitivity curve by measuring the standard deviation of fluxes in 
apertures of 3 pixel diameter and measured the value at the predicted angular separation of Fomalhaut b.
Finally, we noted that the Strehl ratio at the location of Fomalhaut b is degraded due to anisoplanatism.  
We estimated a decrease in sensitivity of 0.75 mag, which corresponds to an isoplanatic angle of 
13$^{\prime\prime}$ at 1.6 $\mu$m.

Calibration of the Gemini $L'$-band data was performed using observations 
of a standard star, HR 9016A, which were obtained in the middle of the observing 
sequence each night.
The unsaturated exposures of the standard were used to derive a photometric zero point and aperture correction for each night.
The data from the different nights were scaled to the same throughput in apertures of 0.3$^{\prime\prime}$ diameter prior to combination.
The upper limit to Fomalhaut b was obtained by measuring the standard deviation of flux in 0.3$^{\prime\prime}$ apertures in an arc along $\pm$45 degrees of the predicted position of Fomalhaut b.
We adopt a factor of 2 in decreased sensitivity due to the estimated Strehl degradation from errors in centroiding the saturated images of Fomalhaut.\\

{\noindent\bf Bolometric luminosity}

In this section and following, we consider various possibilities for
the origin of the detected optical flux.  Here, we assume that the
F814W flux is pure thermal emission from the planet, which consistent
with the model atmosphere from ({\it S7}) where $T_{eff} = 400$ K; g =
46 m s$^{-2}$; and 5$\times$ solar metallicity.  With this effective
temperature, and with a planet radius 1.2 $R_J$, the bolometric flux
at Earth is:
\begin{equation}
F = \biggr({r_p \over d_p}\biggl)^2 ~\sigma_{SB} ~ T_{eff}^4 = 1.86\times10^{-13} ~{\rm erg~ s^{-1}~cm^{-2}}
\end{equation}
where $r_p$ is the the radius of Fomalhaut b, $d_p$ is the heliocentric distance, and $\sigma_{SB}$ is
the Stefan-Boltzmann constant.  As a check, we integrate the flux from the ({\it S7}) high resolution
model spectrum, and obtain:
\begin{equation}
 \int\limits_{6.0\times10^{12}~{\rm Hz}}^{7.5\times10^{14}~{\rm Hz}} F_\nu~ d\nu = 1.80 \times 10^{-13} {\rm erg ~s^{-1}~cm^{-2}}
 \end{equation}
The smaller value is expected because the model is tabulated over a finite frequency range and some
power is missing in the numerical integration.  The corresponding luminosity
is $\sim3\times10^{-7}$ L$_\odot$, which indicates that Fomalhaut b is the faintest known 
object outside of the solar system.\\

{\noindent\bf H$\alpha$ Emission}

Photometry in the F606W filter varies between 25.1 and 24.3 mag (0.36 and 0.75 $\mu$Jy, respectively).
Assuming that the flux is due to a single, narrow emission line, the equivalent line flux would be
$\Delta\nu F_\nu = 0.7 - 1.5 \times 10^{-14} ~{\rm erg ~s^{-1}~cm^{-2}}$.  The fractional luminosity would be
$L_{\rm H_\alpha} / L_{\rm bol} = 0.4 - 0.8 \%$.  The H$\alpha$ emission from
Fomalhaut b would be similar to that suggested for GQ Lup b, where $L_{\rm H_\alpha} / L_{\rm bol} = 2 \%$ ({\it S8}).

The H$\alpha$ emission hypothesis can be tested with an optical spectrum of
Fomalhaut b.  If confirmed, then a key problem is explaining the origin 
of gas around a 200 Myr yr old star (two orders of magnitude
older than GQ Lub b).  Equating the H$\alpha$ luminosity to
accretion luminosity, the accretion rate is 10$^{-11}$ M$_J$ yr$^{-1}$,
or 0.002 M$_J$ over the age of the system (assuming 100\% efficiency).
If we assume that the efficiency is $\sim$1\%, then the total gas accretion is
0.2 M$_J$.\\

{\noindent\bf Dust Cloud Model}

We explore the possibility that Fomalhaut b represents reflected light from an unresolved dust cloud that is not
gravitationally bound and therefore not associated with a planet.  In this scenario
the cloud arises from the stochastic, catastrophic collision of two parent bodies analogous to Kuiper Belt
Objects or short-period comets in the solar system.  The event is improbable
at the location of Fomalhaut b compared to regions closer to the star where
the collision timescales are significantly shorter, or farther from the star where the number density of 
parent bodies is enhanced in order to replenish the visible belt with fresh dust.  

Since Fomalhaut b
appears as a point source in the HST data, the maximum size of the dust cloud
corresponds to the full-width at half-maximum of the PSF, which is 69$\pm$6 mas or
0.53 $\pm$0.05 AU (compared against the
background star shown in Fig. S1, which has FWHM = 68 $\pm$ 4 mas).   A dust cloud could originate from
a catastrophic collision between two planetesimals, but the event must be
recent because even in the absence of stellar radiation pressure and Poynting-Robertson
drag, the different orbital period of a dust grain located at the inner boundary of the cloud (i.e. closest
to Fomalhaut) versus the outer boundary of the cloud would shear the cloud into an arc, ultimately becoming
a ring of material orbiting Fomalhaut.  

The dust cloud will contain a size distribution of grains, though the scattered light images
are predominantly sensitive to grain sizes with $x$ = 2$\pi a$/$\lambda$ $\sim$ 1, where
$a$ is the grain radius.  In our model of a dust cloud we assume a size distribution
with $a_{min} < a < a_{max}$ following a differential size distribution $dn/da = n_o (a/a_o)^{-3.5}$.
We note that due to radiation pressure from Fomalhaut, grains smaller than 3$-$8 $\mu$m (depending
on porosity) are ejected from the system on free-fall timescales ({\it S9}).
We therefore use Mie theory to calculate the apparent magnitude and
scattered light color of a dust cloud with $a_{min}$ = 0.01 $\mu$m and $a_{max}$ = 1000 $\mu$m ($m^{0.01}$
in Table S3) and $a_{min}$ = 8 $\mu$m and $a_{max}$ = 1000 $\mu$m ($m^{8}$
in Table S3).  These values represent two extremes of a fresh dust cloud with small grains still
present within the cloud, and a later epoch where only grains larger than the radiation pressure 
blowout size of $\sim$8 $\mu$m have survived.
We test grains composed of water ice 
(density = 1.0 g cm$^{-3}$; $m_{\rm ice}$ in Table S3) and refractory carbonaceous material 
(density = 2.2 g cm$^{-3}$; {\it S10}; $m_{\rm LG}$ in Table S3).
The results for these two calculations are given in Table S3.  The total grain mass (and hence the total
scattering surface area) is adjusted such
that the integrated light in F814W from the model matches the observations.
In the case of $m_{\rm ice}^{8}$, the total mass is 1.24$\times10^{21}$ g, which
corresponds to the disruption of a 67 km water ice body.  However, the total
grain mass depends strongly on the value selected for $a_{max}$.  Perhaps a
more useful calculation is the minimum grain mass assuming the grain size distribution
is nearly monodisperse and peaks where the scattering efficiency is highest.  For these
optical observations, the scattering efficiency is highest for grains $0.1-0.2~\mu$m in
size, giving a minimum dust mass in the cloud $M_d = 4.1\times10^{18}~ (\rho_g / 1.0~{\rm g ~cm}^{-3})$ g.
Therefore, for water ice, the minimum grain mass is $4.1\times10^{18}$ g, corresponding to
the disruption of a 10 km radius object.  
\\ 

\vspace{0.1cm}
\begin{tabular}{lllllllllllll}
\multicolumn{7}{c}{\bf Table S3:  Dust cloud model for Fomalhaut b} \\
\hline
Filter & $m_\star$ & $m_{\rm Fom-b}$ & $m_{\rm ice}^{0.01}$ & $m_{\rm LG}^{0.01}$& $m_{\rm ice}^{8}$ & $m_{\rm LG}^{8}$\\ 
\hline
F435W & 1.25 & $>$24.7 3-$\sigma$ & 24.37 & 24.59&24.50&24.83\\
F606W & 1.16 & 25.13 $\pm$ 0.09 & 24.46 & 24.57&24.67&24.68\\
F814W & 1.08 & 24.55 $\pm$ 0.13 & 24.55 & 24.55&24.55&24.55\\
\hline
\end{tabular}
\vspace{0.2cm}
\\

\vspace{0.5cm}
\begin{tabular}{lllllllllllll}
\multicolumn{7}{c}{\bf Table S4:  Colors of the dust cloud model} \\
\hline
Filter & $\Delta m_\star$ & $\Delta m_{\rm Fom-b}$ & $\Delta m_{\rm ice}^{0.01}$ & $\Delta m_{\rm LG}^{0.01}$& $\Delta m_{\rm ice}^{8}$ & $\Delta m_{\rm LG}^{8}$\\ 
\hline
F435W - F606W& 0.09 	& $<$0.43 	& -0.09 & 0.02&-0.17&0.15\\
F606W - F814W& 0.08 	& 0.58 		& -0.09 & 0.02&0.12&0.13\\
\hline
\end{tabular}
\vspace{0.5cm}
\\

In Table S4 we give the apparent optical colors.
The dust cloud model explains the Fomalhaut b observations with respect
to the non-detections in $H$ and $L^\prime$.  Otherwise, the dust cloud model is inconsistent
with the Fomalhaut b photometry because:  (1)  A dust cloud should have been detected in the F435W data
(except in the case of $m_{\rm LG}^{8}$ in Table S4),
and (2) The color of a dust cloud is significantly bluer than the observed red color of Fomalhaut b (Table S3).
A third significant problem with the dust cloud model is explaining the F606W
variability observed over two epochs.  The cumulative scattering surface
area would have to drop by 63\% over 1.73 year to account for the 0.5 mag
decrease in optical magnitude.  One possible mechanism is that the
2004 data show Fomalhaut b with a small grain component, but in 2006 the
small grains have dispersed due to radiation pressure.  
Removing all grains $<8 ~\mu$m in size from the first model without
renormalizing the F814W flux results in a brightness
decrease of 2.23 mag in F606W.  Thus an 0.5 mag decrease is possible by tuning
the removal of small grains over 1.73 years.  This scenario demands that
we have observed the cloud at a fortuitous time right after it has been produced,
but before all of the small grains are blown out.   

Taking all four arguments together -- the low probability of the stochastic
collision at the position of Fomalhaut b, the fortuitous timing to explain variability, the non
detection in the F435W filter, and the somewhat discrepant observed optical colors compared
to a model --  the dust cloud hypothesis appears inadequate to explain the observed properties of Fomalhaut b.
 \\

 {\noindent\bf Reflected light from a planet surrounded by an extended dust disk}

 We consider the hypothesis that the Fomalhaut b observations
 are explained by reflected light from a Jovian planet surrounded by a
 large ring system. First, we consider reflected light from the planet
 alone.  The flux received at Earth from
 the star (Fomalhaut) is:

 \begin{equation}
f_\star = {L_\star \over 4\pi~D^2}  = {6.34\times10^{27} \over 4\pi~(2.379\times10^{17})^2} = 8.914\times10^{-9}~{\rm Wm^{-2}}
\end{equation}

\noindent
where $L_\star$ is the stellar luminosity in watts (W; {\it S12}), and D is the heliocentric
distance (7.688 pc = $2.379\times10^{17}$ m).  The stellar flux received by a planet
at $d$=115 AU radius from Fomalhaut is:

 \begin{equation}
f_o = {L_\star \over 4\pi~d^2}  = {6.34\times10^{27} \over 4\pi~(115 \times1.5\times10^{11})^2}= 1.70~{\rm Wm^{-2}}
\end{equation}

\noindent The flux received at earth:

 \begin{equation}
f_p = {f_o \over 4\pi~D^2}  = {\sigma_p Q_s \times 1.70~ {\rm Wm^{-2}} \over 4\pi~(2.379\times10^{17})^2} = \sigma_p Q_s \times 2.390\times10^{-36}~{\rm Wm^{-2}}
\end{equation}

\noindent where $\sigma_p$ [m$^2$] is the projected geometric surface area of the planet and $Q_s$ is the scattering efficiency, such as the product of the geometric albedo and the scattering phase function at the observed phase.
It is useful to consider these values as a relative contrast in apparent magnitude:

 \begin{equation}
\Delta m = m_p - m_\star = -2.5~ {\rm log} \biggl({f_p \over f_\star}\biggr) =  -2.5~ {\rm log} \biggl({\sigma_p Q_s\times 2.39\times10^{-36} \over 8.91\times10^{-9}}\biggr) = -2.5~ {\rm log}(\sigma_p Q_s) + 69
\end{equation}

\noindent The $V$ band (F606W) apparent magnitude of Fomalhaut is
 $m_V$ = 1.2 mag, giving $m_p = -2.5~ {\rm log} (\sigma_p Q_s)$ +
 70.2.  If we ignore the reduction in brightness due to viewing phase,
 the geometric cross section of a 1.2 $R_J$ planet is $\sigma_p
 = \pi \times (1.2 \times 7.15\times10^7 {\rm m})^2 =
 2.31\times10^{16}$ m$^2$.  For $Q_s = 0.5$, the apparent magnitude of
 this planet $m_p$ = 30.0 mag.  Thus, the observed
 apparent magnitude of Fomalhaut b at optical wavelengths (Table S2)
 is $\sim$5 mag
 brighter than light reflected from a Jupiter.

Now we consider that the planet is surrounded by dust grains analogous
to circumplanetary rings.  Since
this is a flattened disk, the geometric cross section scales as
cos($i$), where $i$ is the inclination to the line of sight ($i=0^{\rm
o}$ is a face-on orientation).  Consider that the main, optically thick rings
of Saturn extend out to Saturn's Roche radius, or about 2 planetary radii.
If Fomalhaut b also harbored a ring system extending to its Roche radius,
then $\sigma_{p}$ would be replaced by the ring cross section $\sigma_{pr} \sim \sigma_p \times 2^2 \times \cos (66^o) \sim 1.6 \sigma_p$.
Assuming again $Q_s$ = 0.5, the apparent magnitude of the system would be
$m_{pr}$ = 29.5 mag, or about 4.5 mag too faint compared to what
is actually observed.

To make up for this shortcoming,
the scattering surface area of the planet+ring system
would have to increase by yet another factor of $\sim$60.
This would correspond to about $\sim$16 planetary radii. If
the effective albedo of the ring particles is closer to 0.1---and in fact
outer solar system albedoes are typically this low---then the rings
must extend to $\sim$35 planetary radii. An optically thick
ring system that is 16--35 planetary radii large is better described
as a protosatellite, circumplanetary disk. For example, the outermost
Galilean satellite of Jupiter, Callisto, has a planetocentric distance
of about $\sim$27 Jupiter radii. Regular satellites have prograde motion that
indicates formation $in~situ$ around the planet.  

Though in some respects this is similar to the pure
dust cloud model, the planet+disk hypothesis has several advantages:
(1) The presence of a planet allows for a wider range of physical
phenomena to account for the F606W variability, such as the H$\alpha$
hypothesis.  (2) The location of Fomalhaut b just inside the dust belt
is consistent with the predicted location of a planet gravitationally
sculpting the belt's inner edge. (3)  The existence of a planet
permits a system of dust to be spatially confined rather than
dispersed due to shearing or radiation pressure.\\

{\noindent\bf SOM References}

\begin{enumerate}

\item H. C. Ford, {\it et al., S.P.I.E.} {\bf 4584}, 81 (2003).
\item C. Marois,  {\it et al.,  Astrophys. J.}  {\bf 641}, 556 (2006).
\item M. P. Fitzgerald, P.G. Kalas, G. Duchene, C. Pinte,  \& J.R. Graham  {\it et al.,  Astrophys. J.}  {\bf 670}, 536 (2007).

\item D. Lafreni\`{e}re {\it et al.,  Astrophys. J.}  {\bf 670}, 1367 (2007).
\item R. P. van der Marel, J. Anderson, C. Cox, V. Kozhurina-Platais, M. Lallo, E. Nelan, Instrument Science
Report ACS 2007-007 (Space Telescope Science Institute; Baltimore), (2007).
\item Sirianni, M.,  {\it et al.,  PASP}  {\bf 117}, 1049 (2005).

\item J.J. Fortney,  {\it et al.,  Astrophys. J.}  {\bf 683}, 1104 (2008).
\item C. Marois, B. Macintosh, T. Barman, {\it Astrophys. J.} {\bf 654}, L151 (2007).
\item P. Artymowicz \& M. Clampin {\it Astrophys. J.} {\bf 490}, 863 (1997).

\item A. Li, J. M. Greenberg, {\it Astron. Astrophys} {\bf 323}, 566 (1997).
\item J. Davis, {\it et al.,  Astron. Nachr.}  {\bf 326}, 25 (2005).

\end{enumerate}

\bibliography{scibib}

\bibliographystyle{Science}

\clearpage

\begin{figure}[htbp]
\begin{center}
\includegraphics[width=\textwidth]{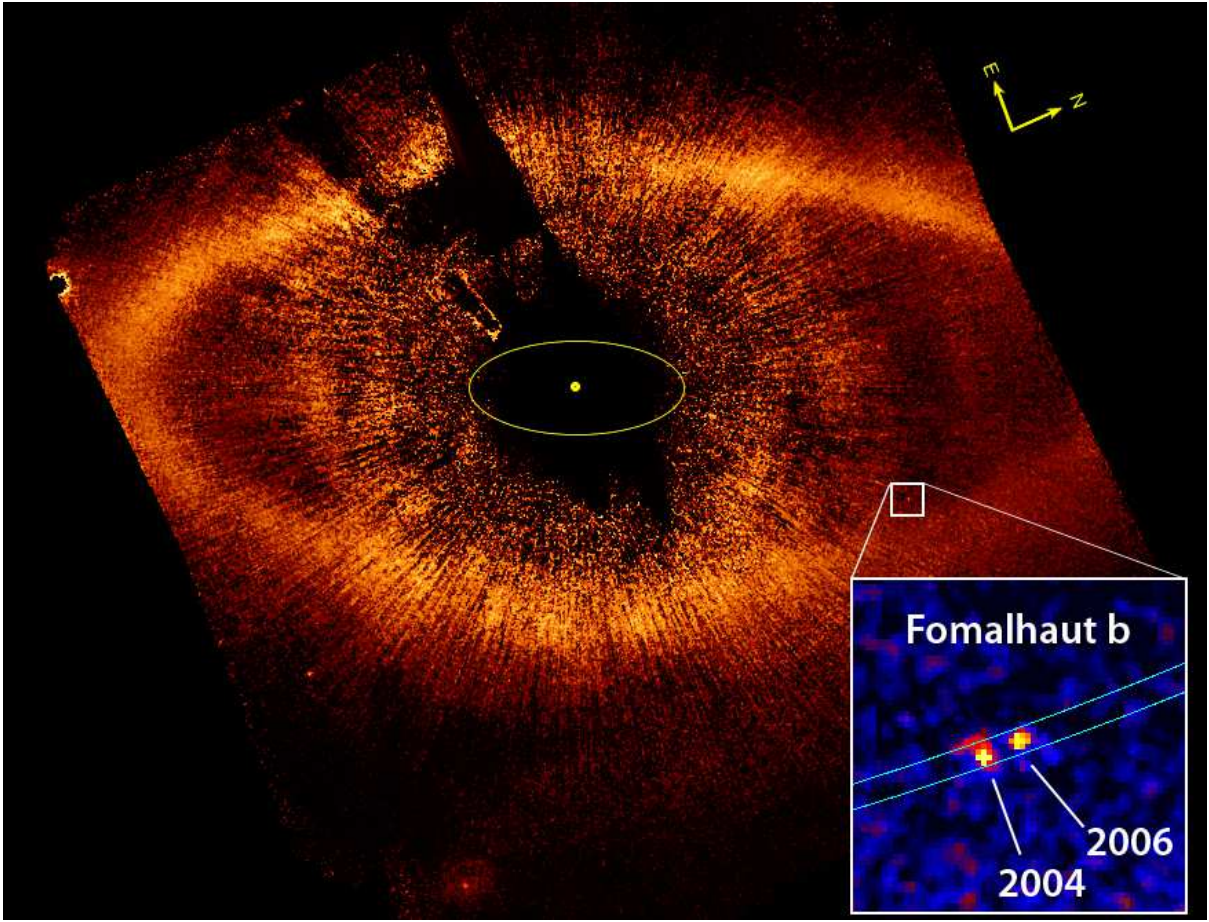}
\end{center}
\caption{HST coronagraphic image of Fomalhaut at 0.6 $\mu$m showing the location 
of Fomalhaut b (white square) 12.7$^{\prime\prime}$ radius from the star and
just within the inner boundary of the dust belt.  All the other apparent 
objects in the field are either background stars and galaxies or false-positives.  
The fainter lower  half of the dust belt lies behind the sky plane.  
To obtain an orientation with north up and east left, this figure should be
rotated 66.0$^{\rm o}$ counterclockwise.  The yellow circle marks the location of the star behind
the occulting spot.  The yellow ellipse has a semimajor axis of 30 AU at Fomalhaut (3.9$^{\prime\prime}$)
that corresponds to the orbit of Neptune in our solar system.  The inset is a composite
image showing the location of Fomalhaut b in 2004 and 2006 relative to Fomalhaut.
Bounding Fomalhaut b are two elliptical annuli
that are identical to those shown for Fomalhaut's dust belt ({\it 6}), except that here the inner
and outer annuli have semimajor axes of 114.2 and 115.9 AU, respectively.  The motion of
Fomalhaut b therefore appears to be nested within the dust belt.
}
\end{figure}

\begin{figure}[htbp]
\begin{center}
\includegraphics[width=3.3in]{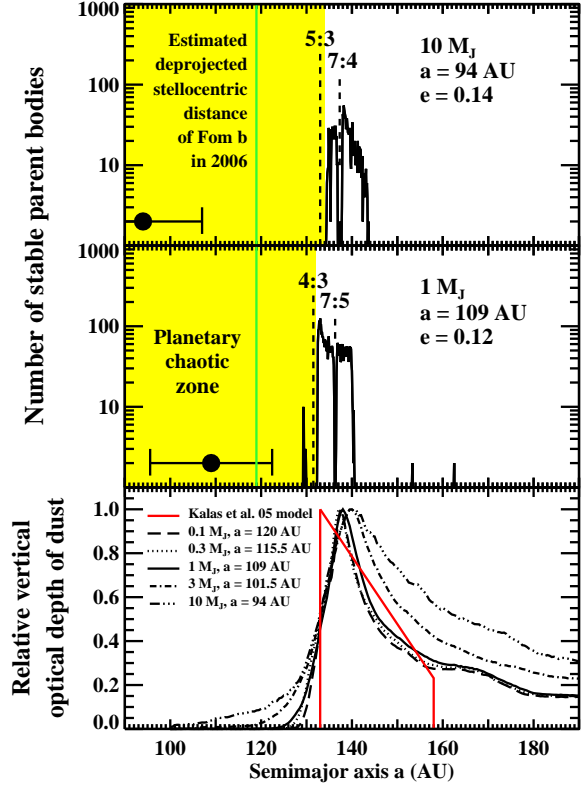}
\caption{Dynamical models of how Fomalhaut b gravitationally sculpts
the belt [see also ({\it 14})]. {\it Top two panels}: Histograms of time-averaged semimajor
axes of parent bodies that survive 100-Myr-long integrations with
Fomalhaut b, whose parameters are chosen to reproduce the belt's inner
edge at 133 AU and ellipticity of 0.11.  Parent bodies are evacuated
from Fomalhaut b's chaotic zone (yellow region).  Gaps open at the planet's
resonances, akin to the solar system's Kirkwood gaps. Black circles
and bars mark the range of stellocentric distances spanned by the
model orbits for Fomalhaut b. The apocentric distance for $10 ~M_{\rm J}$ is
inconsistent with the observed stellocentric distance of Fomalhaut b (green
line). The $1~ M_{\rm J}$ model is consistent. {\it Bottom panel}:
Vertical optical depth profiles of dust generated from parent bodies.
The planet orbit is tuned so that the optical depth is at half maximum
at 133 AU, the location of the inner edge of the 
scattered light model from ({\it 6}) (red curve), which itself is an idealized and
non-unique fit to the HST data. While the dynamical and scattered
light models do not agree perfectly, lower planet masses are still
inferred because they do not produce broad tails of emission at $a
\gtrsim 140$ AU.  At $a\gtrsim 160$ AU, the HST data are too
uncertain to constrain any model.}
\end{center}
\end{figure}

\begin{figure}[htbp]
\begin{center}
\includegraphics[width=\textwidth]{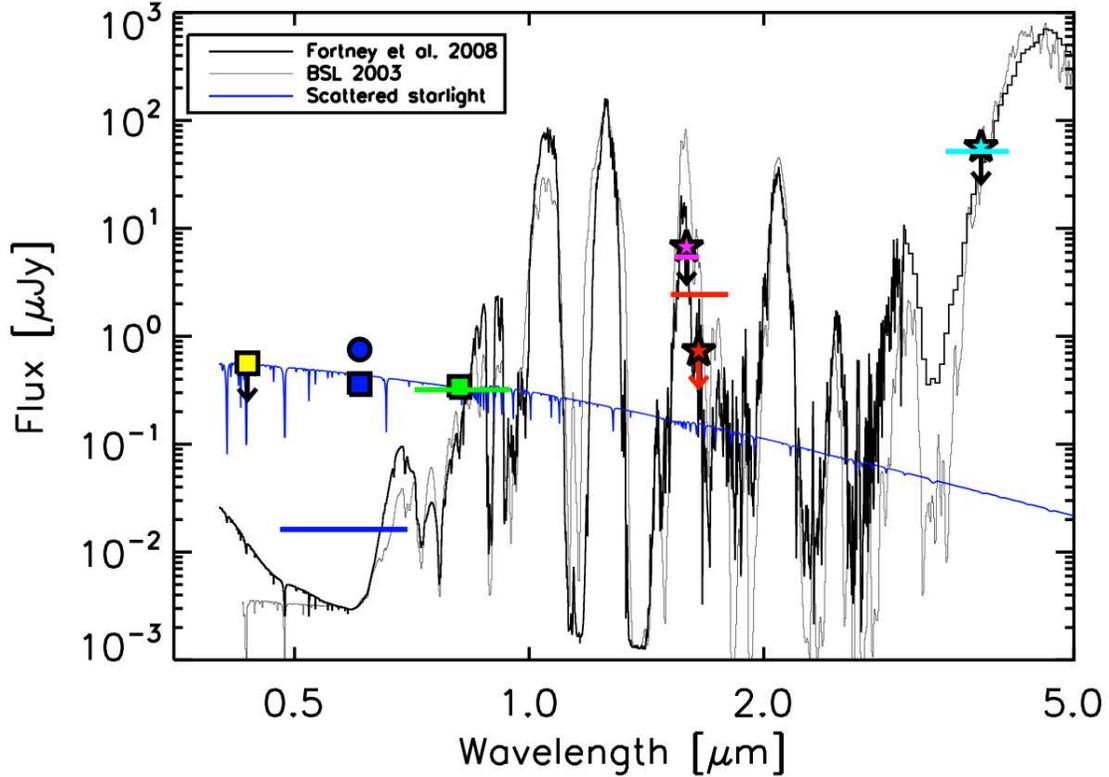}
\caption{Photometry on Fomalhaut b shows the F435W 3-$\sigma$ upper
  limit (yellow square), two F606W measurements (blue square=2006,
  blue circle=2004), the F814W photometry (green square), 3-$\sigma$
  upper limits for Keck observations in the CH$_4$ passband (purple
  solid star) and the H band (red solid star), and a 3-$\sigma$ upper
  limits for Gemini observations at L$^\prime$ (light blue star).
  This is a log-log plot.  If we first assume that the F606W
  variability is due to H$\alpha$ emission and the F814W detection is
  due to planet thermal emission, we then proceed to fit a planet
  atmosphere model from ({\it 15}) to the F814W flux.  The heavy solid
  line represents that planet atmosphere model smoothed to R=1200 with
  planet radius 1.2 R$_J$, gravity 46 m s$^{-2}$, and T=400K (roughly
  1-3 M$_J$ at 200 Myr).  The horizontal colored lines mark the
  equivalent broad-band flux found by integrating the model spectrum
  over the instrumental passband.  Other models from ({\it 16}) give a
  similar spectrum (light solid line), though a factor of 3 - 4
  brighter in CH$_4$ and H band.  The model predicts that the planet
  candidate should have been detected with Keck in the H band, though
  this prediction is only a factor of a few above our limit. The
  discrepancy could arise from uncertainties in the model atmosphere
  (which has never been tested against observation), or from the
  possibility that the F606W and F814W detections include stellar
  light reflected from a circumplanetary dust disk or ring system.
  The solid blue line intersecting the optical data represents light
  reflected from a circumplanetary disk with radius 20 R$_J$, a
  constant albedo of 0.4, and with stellar properties adopted from
  ({\it 22}).}
\end{center}
\end{figure}

\begin{figure}[htbp]
\begin{center}
\includegraphics[width=\textwidth]{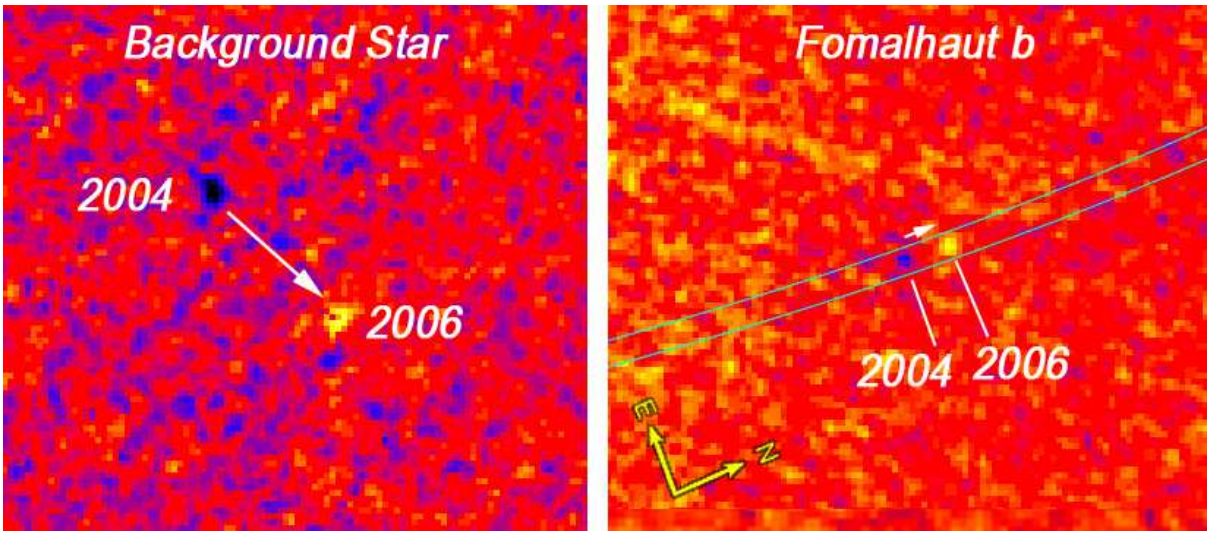}
\end{center}
\caption{{\bf Fig. S1}:  Two enlarged sub-regions (at the same scale) from Figure 1 centered on Fomalhaut b and a background star (located
at the 8 o'clock position relative to Fomalhaut in Fig. 1, just outside the dust belt).  
We show relative motion by registering the 2004 and 2006 data to Fomalhaut and producing the difference image.  
Background objects are easily distinguished from the planet candidate in terms of the 
magnitude (0.7 arcsecond) and direction of their motion.  In 2004, Fomalhaut b 
is detected at separation $\rho$ = 12.61$^{\prime\prime}$ and position angle, PA = 316.86$^{\rm o}$ relative to Fomalhaut.  
In 2006, Fomalhaut b is at  $\rho$ = 12.72$^{\prime\prime}$ and position angle, PA = 317.49$^{\rm o}$
(recall that the orientation shown here is rotated 66.0$^{\rm o}$ clockwise
from one that gives north up and east left).  
}
\end{figure}

\end{document}